\documentclass{revtex4-2}
\usepackage{siunitx}
\usepackage{graphicx}

\begin{document}

\title{Healing Gradient Degradation in Nb\textsubscript{3}Sn SRF Cavities Using a Recoating Method}
\author{Eric Viklund}
\email{ericviklund2023@u.northwestern.edu}
\affiliation{Department of Materials Science and Engineering, Northwestern University}
\affiliation{Fermi National Accelerator Laboratory}
\author{David N. Seidman}
\affiliation{Department of Materials Science and Engineering, Northwestern University}
\author{Sam Posen}
\email{sposen@fnal.gov}
\affiliation{Fermi National Accelerator Laboratory}
\author{Brad M. Tennis}
\affiliation{Fermi National Accelerator Laboratory}
\author{Grigory Eremeev}
\affiliation{Fermi National Accelerator Laboratory}

\date{\today}

\begin{abstract}

    Despite having advantageous superconducting properties, Nb\textsubscript{3}Sn superconducting radiofrequency (SRF) cavities still have practical challenges compared to Nb SRF cavities due to the brittle nature of Nb\textsubscript{3}Sn. Performance degradation can occur when a Nb\textsubscript{3}Sn SRF cavity experiences mechanical stresses such as during handling and tuning of the cavity. In this study, we present a potential treatment for SRF cavities that have experienced stress induced performance degradation that involves a recoating procedure. The degraded cavity is coated with a small amount of Sn using a single step vapor-diffusion methodology. Using this approach we can recover a significant portion of the lost performance of a Nb\textsubscript{3}Sn SRF cavity.

\end{abstract}

\maketitle

\section{Introduction}
\label{sec:Introduction}

Niobium cavities have been extensively studied and treatments have been developed to optimize the accelerating gradient and quality factor~\cite{10.1063/1.4866013, 10.1063/1.4960801, 10.1063/5.0059464, 10.1063/5.0063379}. The performance of niobium SRF cavities is limited by the material properties of niobium (Nb). A promising alternative to Nb is Nb\textsubscript{3}Sn. There exists a large body of research on creating high performance Nb\textsubscript{3}Sn superconducting radiofrequency (SRF) cavities\cite{10.1063/1.4913617, 10.1063/1.4913247}. Desirable superconducting properties, such as higher superconducting transition temperature (T\textsubscript{c}) and a higher superheating magnetic field (H\textsubscript{sh})\cite{liarte2017theoretical, catelani2008temperature, lin2012effect, kubo2020superfluid}, make Nb\textsubscript{3}Sn an attractive material for SRF applications. The material properties of Nb\textsubscript{3}Sn make,however, it difficult to work with. 

The brittleness of Nb\textsubscript{3}Sn introduces new challenges to the cavity manufacturing process. Nb\textsubscript{3}Sn must be deposited as a thin film on a bulk cavity substrate\cite{posen2017nb3sn, pudasaini2019growth, porter2018update}. Because of the thin and brittle film, Nb\textsubscript{3}Sn cavities are highly susceptible to mechanical stress. Nb\textsubscript{3}Sn cavity performance is known to degrade when stresses are applied to the cavity\cite{eremeev2023preservation, eremeev:srf2019-mop015}. This degradation is assumed to be caused by cracks in the brittle Nb\textsubscript{3}Sn film caused by deformation of a cavity during processing such as tuning or assembly. Cavities that suffer from degradation are typically stripped and recoated with a new Nb\textsubscript{3}Sn film, which is a time-consuming and expensive process.

In this current study we explore a new procedure to heal Nb\textsubscript{3}Sn cavities whose performance has been degraded by deformation. This procedure utilizes a short Nb\textsubscript{3}Sn recoating to attempt to heal cracks that have formed in the cavity without the need to remove the original film. This procedure was developed to restore the performance of a Nb\textsubscript{3}Sn cavity which has undergone centrifugal barrel polishing\cite{viklund2024improving}. The performance decrease measured on a polished cavity is like the above mentioned case of deformation-induced degradation. When employing this recoating procedure to a degraded cavity, we can recover a large portion of the performance with a simple furnace treatment. This discovery provides a valuable method for recovering degraded cavities without lengthy reprocessing which avoid subsequent thinning and frequency shifts.

\section{Experiment}
\label{sec:Experiment}

This study is performed on a Nb\textsubscript{3}Sn, \qty{1.3}{\giga\hertz} cavity coated using a high-temperature nucleation step to create a Nb\textsubscript{3}Sn film with low surface roughness. An in-depth analysis of this cavity coating and the initial performance of the cavity can be found in reference~\cite{posen2021advances}. 

After initial testing, the cavity was transported to Cornell, after which performance decreased. The cavity was then returned back to FNAL for additional testing, which confirmed the performance degradation. We suggest that the degradation was caused by stresses applied to the cavity during transport, which led to the formation of cracks. This type of performance degradation has previously been observed during assembly of Nb\textsubscript{3}Sn cavities~\cite{eremeev2023preservation}, and when tuning Nb\textsubscript{3}Sn cavities at room temperature~\cite{eremeev:srf2019-mop015}. In these cases, stresses applied to the cavity were suggested to be the main cause of the degradation. This indicates that the stresses must be carefully controlled when handling Nb\textsubscript{3}Sn cavities otherwise cracks can form in the Nb\textsubscript{3}Sn film.

To heal the cracks causing the performance degradation, we apply a recoating procedure. During this recoating procedure, the cavity was heated to \qty{1000}{\degreeCelsius} and exposed to Sn vapor for \qty{1}{\hour}. Sn vapor was provided by \qty{0.85}{\gram} of Sn heated to \qty{1250}{\degreeCelsius}. The reasoning behind these parameters is that only a small amount of Sn is necessary to fill the microscopic cracks in the film. Applying too much Sn causes the film to become too thick and negatively impacts the surface roughness of the film. During the coating process only a small fraction of the Sn evaporated leaving behind a large amount of the initial Sn still in the crucible. 

\section{Results}
\label{sec:Results}

\begin{figure}[h]%
    \centering%
    \includegraphics[width=1.0\columnwidth]{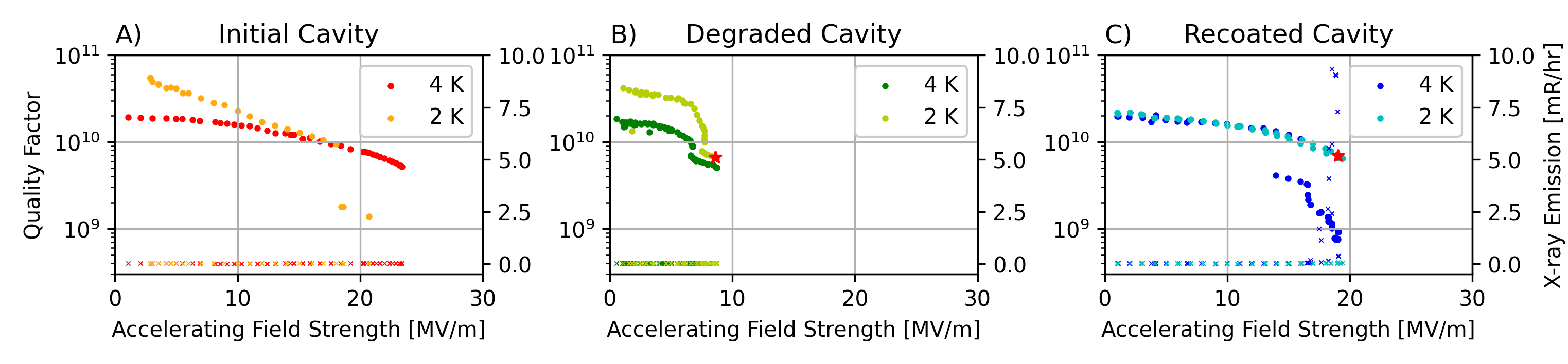}%
    \caption{The quality factor, indicated by dots, and X-ray emissions, indicated by x, versus the accelerating gradient of the cavity after the initial coating (A), after the degradation (B), and after the recoating (C). The quality factor and accelerating gradient of the T-maps in figure~{\protect\ref{fig:TMAP}} are indicated by a red star.}%
    \label{fig:VTS}%
\end{figure}

\begin{figure}[h]%
    \centering%
    \includegraphics{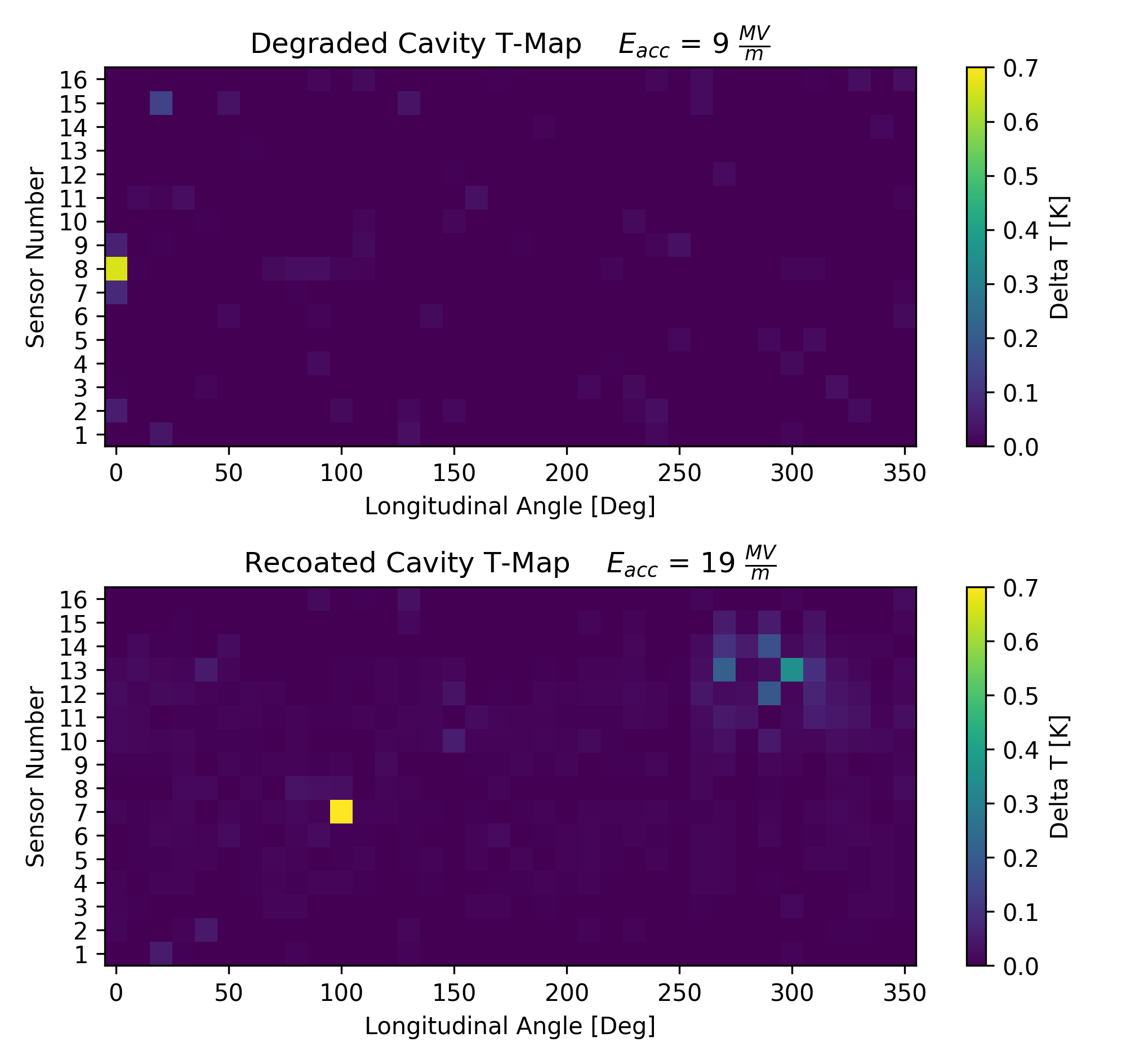}%
    \caption{Temperature maps of a cavity's surface prior to quench as measured before (top) and after (bottom) the recoating is applied. The temperature of the hot spot near the equator of the recoated cavity exceeds the maximum value of the color bar and achieves a maximum value at \qty{3}{\kelvin}}%
    \label{fig:TMAP}%
\end{figure}

After initially coating the cavity it achieves a peak accelerating field of \qty{24}{\mega\volt\per\meter} and a maximum Q of \num{2e10} at \qty{4}{\kelvin}. The peak accelerating gradient after degradation is \qty{8}{\mega\volt\per\meter} and a maximum Q of \num{1e10} at \qty{4}{\kelvin}. The cavity displays a decrease in the quality factor at around \qty{6}{\mega\volt\per\meter} before a quench. A similar decrease in the quality factor is observed for other cavities affected by performance degradation\cite{eremeev2023preservation,eremeev:srf2019-mop015} and in Nb\textsubscript{3}Sn cavities treated with centrifugal barrel polishing\cite{viklund2024improving}. Temperature mapping is displayed in figure~\ref{fig:TMAP} and is utilized to locate the quench source responsible for the performance degradation. A single hot spot on the equator of the cavity is present. Visual inspection of the cavity did not display visible defects near the quench location.

After the recoating procedure is applied the cavity's performance increases. The peak accelerating gradient increases to \qty{19}{\mega\volt\per\meter} and the quality factor is \num{1e10} at \qty{4}{\kelvin}. Temperature mapping of the cavity after recoating demonstrates that the initial hot spot is healed with no detectable heating from that area. This indicates that the defect causing the performance degradation is repaired by recoating. At higher gradients another small hotspot appears in a new location close to the equator. Additionally, there is also a larger hot spot closer to the iris, which appears just before the cavity quench. This additional hotspot is accompanied by a large increase in x-ray emissions, figure~\ref{fig:VTS}, which may indicate that heating is caused by multipacting. Higher accelerating gradients are not attainable even after \qty{5}{\hour} of processing. The cavity did not display any signs of multipacting during the \qty{2}{\kelvin} measurement.

\section*{Discussion}
\label{sec:Discussion}

Recoating a damaged Nb\textsubscript{3}Sn cavity can have a major impact on its performance. The mechanism for this change is heretofor unknown. We propose two possible mechanisms for the recoating process, which heals the Nb\textsubscript{3}Sn film. The first mechanism is by filling the cracks. When a cavity is exposed to Sn, a thin layer of liquid Sn coats the surface which thereby fills the cracks. Nb\textsubscript{3}Sn is then formed in the cracks by diffusion of Nb into the liquid Sn. The diffusion rate of Nb is relatively slow compared to the diffusion of Sn, however if the cracks are small, less than a few \qty{100}{nm}, Nb may have sufficient time to diffuse into the crack. The second mechanism involves the diffusion of Sn into the Nb substrate through the crack. If the crack penetrates the Nb\textsubscript{3}Sn film, Sn can diffuse into the crack and react with the Nb substrate creating a region of new Nb\textsubscript{3}Sn. This new region acts as a bridge for electrical currents to flow through the film and prevents current from flowing through the Nb substrate, which has a higher resistivity than does Nb\textsubscript{3}Sn. This is an example of self healing~\cite{Sloof2007}.

\section{Conclusion}
\label{sec:Conclusion}

Using a low temperature (\qty{1000}{\degreeCelsius}), short duration (\qty{1}{\hour}) Sn recoating process, we are able to heal a degraded Nb\textsubscript{3}Sn cavity that suffered damage during transportation. The recoating process improved the maximum gradient of the cavity from \qty{8}{\mega\volt\per\meter} to \qty{19}{\mega\volt\per\meter}, which is close to the initial performance of the cavity of \qty{24}{\mega\volt\per\meter}. Temperature mapping measurements of the cavity demonstrate that a single hot spot on the equator of the cavity was responsible for the performance degradation. After the recoating process this defect becomes healed leading to less heating and a higher maximum electric field gradient. Ultimately, the performance is limited by a second hot spot.

This discovery provides a new approach which applies to similarly degraded SRF cavities to recover their performances. This approach saves time and money which would otherwise be spent removing the Nb\textsubscript{3}Sn coating and then applying a new coating. This self healing process makes Nb\textsubscript{3}Sn cavities more viable for real-world accelerator applications by reducing their manufacturing costs.

\section{Acknowledgements}

This manuscript has been authored by Fermi Research Alliance, LLC under Contract No. DE-AC02-07CH11359 with the U.S. Department of Energy, Office of Science, Office of High Energy Physics.

\bibliographystyle{plain}
\bibliography{bib}

\end{document}